\documentclass{aa}  

\usepackage{graphicx}
\usepackage{txfonts}
\usepackage{lipsum}
\usepackage{xcolor}
\usepackage[hidelinks,colorlinks=true,linkcolor=blue,citecolor=blue]{hyperref}
\usepackage{subcaption}         
\usepackage{lscape}             
\usepackage{placeins}           
\usepackage{ulem}  
\usepackage{amsmath}

\begin{document}
   \title{A thick-shell formalism for pulsar wind nebulae based on energy conservation}


   \author{N. Bucciantini\inst{1,2,3}\fnmsep\thanks{email: niccolo.bucciantini@inaf.it}
        \and B. Olmi\inst{1}
        }

\institute{
INAF, Osservatorio Astrofisico di Arcetri, Largo E. Fermi 5, I-50125 Firenze, Italy
\and
Dipartimento di Fisica e Astronomia, Universit\`a di Firenze, Largo E. Fermi 2, I-50125 Firenze, Italy
\and
INFN , Sezione di Firenze, Via G. Sansone 1, I-50019 Sesto Fiorentino (FI), Italy
}

   \date{Received XX XX, 20XX}

 
  \abstract
   {One of the most powerful approach to model the structural and spectral evolution of wind bubbles and pulsar wind nebulae in particular is based on the so called \textit{one-zone thin-shell} formalism. By solving a  set of simple equations one can relate the spectral properties of these systems to the physical properties of the pulsar and supernova remnant. Due to its predictive power, this approach has been widely used both in fitting existing objects, and in population synthesis. However there are some well known inconsistencies when applied to the Crab nebula, which have never been fully accounted for, but whose implications cast serious doubts on its overall reliability.}
   {The introduction of a new and more flexible formalism based on energy conservation that provides a way to reconcile the observed structural and spectral properties of the Crab nebula, solving the known inconsistencies, as well as to model in a more realistic and physically appropriate way the evolution of wind bubbles. } 
   {The equations of the formalism are presented and discussed in details, together with both simplified solutions, and more complex ones including radiation losses. 
   Implications for the modeling of wind bubbles, and pulsar wind nebulae, are illustrated and discussed. Moreover, we introduce a new high-order upwind-implicit scheme for the evolution of the particle spectrum that ensures high-accuracy in energy conservation, and we include an algebraic-vectorizable approach for self-sychrotron Compton in the full Klein-Nishina regime that avoids time consuming interpolations or integrations. }
 {We reproduced the Crab nebula structural properties and spectrum, accounting for the difference in the convergence age between the optical filaments and the radio bubble, removing the known inconsistencies.}
   {The spectral accuracy of our new approach is comparable with the standard one, but it proves superior in its ability to reproduce structural properties, and to account for geometrical effects as  the development of a thick layer of mixed material and the lack of an efficient coupling between the wind bubble and the surrounding medium. }

   \keywords{radiation mechanisms: non-thermal -- pulsar: general -- method: numerical -- ISM: supernova remnants -- supernovae: individual: Crab nebula
               }

   \maketitle

\section{Introduction}
\label{sec:intro}
The one-zone thin-shell (1ZTS) formalism is one of the most used approach for the study of the spectral and dynamical evolution of wind bubbles, and has been successfully applied in the field of pulsar wind nebulae (PWNe) for a long time \citep{Ostriker_Gunn71a,Reynolds_Chevalier84a,Gelfand_Slane+09a,Bucciantini_Arons+11a,Martin_Torres+12a}. The approach is not without its limitations, especially when applied to old systems \citep{Bandiera_Bucciantini+20a,Bandiera_Bucciantini+23a}, and indeed an alternative formalism that still preserve many of the advantages of the old ones, while providing a consistent description of the interaction of the PWN with the supernova remnant (SNR), has been recently put forward in \citet{Bandiera_Bucciantini+23b}. However the lack of viable alternatives (multidimensional fluid models are computationally prohibitive)  makes this technique the only one currently available for spectral studies of both young and old systems, especially in population oriented studies \citep{Fiori_Olmi+22a,DeSarkar26a}.\\
\\
One of the main limitation of all existing  thin-shell models (but in general of all existing 1D models), is what we can call the \textit{complete-coupling} of the PWN (or more generally the wind bubble) with the SNR (the ambient medium). By complete-coupling we mean that the SNR material and the PWN are sharply separated by a contact discontinuity (CD), and that the PWN sweeps and keeps outside of this CD all matter it encounters. As such, existing approaches do not allow to treat the development of a thick filamentary layer of swept-up ejecta, or any form of mixing, which is well known to be present \citep{Hestewr-Strone+99a,Fesen_Rudie+08a,Ma_Ng+10a,Meyer_Torres25a}. This is due to the fact that the canonical equations of the 1ZTS models are based on point-wise momentum conservation across the thin-shell itself \citep[e.g.][]{Bandiera_Bucciantini+20a}.  Therefore this limitation is rooted at the very base of these approaches. This also prevents the possibility for the PWN shape to differ from a spherical one. While the thin-shell equations can be formulated in order to handle aspherical shapes \citep{Giuliani82a}, and models of axisymmetric wind bubbles adopting them have been presented \citep{Buciantini_Quataert+07a}, they tend to be unstable and have never been used for PWNe.  Moreover 1ZTS models do not allow to investigate cases where the PWN penetrates through the ejecta, instead of just pushing against them. \\
\\
Here we want to introduce a new formalism for the evolution of PWNe within the ejecta of SNRs, based on energy conservation. Energy is a scalar quantity for which global conservation holds, such that changes in the energetic content of one component (i.e. the PWN) must be balanced by equivalent changes in the energetic content of another (i.e. the swept up ejecta). Momentum is a vector quantity, but in a centrally symmetric configuration, as the case for a PWN-SNR system, the global integrated momentum of any component is zero by definition, and in this sense global momentum conservation is useless. If one assumes spherical symmetry, it is in principle possible to frame the problem in terms of radial-momentum conservation. However there are in this case two main issues: what is the radial momentum content of the expanding PWN? How does momentum transfer from one component to another? The answer to the first question requires the knowledge of the internal velocity structure of the PWN, the second of the stress-coupling with the ejecta, none of which is known. This is where a formalism based on energy conservation proves superior. In the following we will refer to the formalism based on energy conservation as E-formalism, and the one on momentum conservation as M-formalism.\\
\\
We will show that our E-formalism not only allows for a vastly larger degree of flexibility in modeling PWNe, including effects not otherwise accountable, while retaining all the computational advantages of the canonical M-formalism, but that its application might help to address some long standing issues in the modeling of the Crab nebula, that have never received a due attention, given the role of this object as a test-bed for PWNe in general. 
In particular we will focus on three related issues: (i) the swept-up mass measured in the filaments of the Crab nebula; (ii) the different expansion rate of the filaments versus the radio nebula; (iii) the lack of any detectable low density halo surrounding it. Estimates of the mass in the filaments have gone up from $2.1~M_\odot$ \citep{Henry_MacAlpine82A} to $4.6~M_\odot$ \citep{Fesen_Shull+97A} and the more recent value of  $7.0~M_\odot$ \citep{Owen_Barlow15A}. 
On the other hand the existing 1ZTS models that fit the Crab nebula spectrum predict that the ejecta mass swept-up by the PWN as it expands (the mass that should correspond to the one in the observed filaments) should be much smaller (by about one order of magnitude): $\sim0.5~M_\odot$. Under complete-coupling, the swept-up mass is bound to move with an expansion rate defined as the ratio of the velocity times the age over the radius (in practice a measure of the convergence date - the inferred date of the explosion based on ballistic motion) that is the same as the one of the non-thermal PWN. However, recent measures show that the two differ by about 15\%, with the non-thermal PWN being faster \citep{Bietenholz_Nugent15A}. 
All the existing 1ZTS models of the Crab nebula \citep{,Gelfand_Slane+09a,Bucciantini_Arons+11a,Martin_Torres+12a} predict that it should be still expanding well within the central uniform density region of the ejecta, but the attempts to constrain the outer density distribution suggest a much steeper density profile \citep{Fesen_Shull+97A,Sankrit_Hester97a,Lundqvist_Tziamtzis12a}. \\
\\
We will show that such discrepancies cannot be reconciled in the classical 1ZTS formalism by changing either the SNR energy or ejecta mass, and that they are ultimately tied to complete-coupling. Our E-formalism is not constrained by complete-coupling, and can accommodate the correct results. This, however, cast doubts both on the overall reliability of canonical 1ZTS models in the way they are used, and on the interpretation of the origin of the observed filaments themselves. A point that in our opinion need to be clarified.\\
\\
This manuscript is structured as follows: 
in Sect.~\ref{sec:eqns} we introduce the equations of the E-formalism. In Sect.~\ref{sec:sefsim} we derive a self-similar solution that can be used to gain some insight into the properties of the model, while in Sect.~\ref{sec;crab} we discuss more in detail the case of the Crab nebula, and illustrate how the E-formalism allows us to reproduce consistently both the structural and spectral properties of this system. Finally in Sect.~\ref{sec:conclusions} we summarize our conclusions.

\section{The equations}
\label{sec:eqns}
Let us write down the set of full general equations for the E-formalism. We begin with the evolutionary laws:
\begin{align}
    \frac{{\rm d}E_{\rm pwn}^{\rm p}(t)}{{\rm d} t} &= \eta_{\rm p}L_{\rm psr}(t) - Q_{\rm loss}(t)\,,\label{eq:e1}\\
    \frac{{\rm d}E_{\rm pwn}^{\rm m}(t)}{{\rm d} t} &= \eta_{\rm b}L_{\rm psr}(t) - P_{\rm pwn}^{\rm m}(t)\frac{{\rm d}\mathcal{V}_{\rm pwn}(t)}{{\rm d} t}\,,\label{eq:e2}\\
    \frac{{\rm d}W_{\rm pwn}(t)}{{\rm d} t} &= [ P_{\rm pwn}^{\rm p}(t)+P_{\rm pwn}^{\rm m}(t)]\frac{{\rm d}\mathcal{V}_{\rm pwn}(t)}{{\rm d} t}\,,\label{eq:e3}\\
    E_{\rm fil}(t)&= W_{\rm pwn}(t) +E_{\rm ini}(t)\,,\label{eq:e4}
\end{align}
where $E_{\rm pwn}^{\rm p}(t)$ is the plasma (particles) energy of the PWN, while $E_{\rm pwn}^{\rm m}(t)$ is its magnetic energy. $\eta_{\rm p}$ is the fraction of the PSR luminosity $L_{\rm psr}(t) $ that goes into particles, and $\eta_{\rm b}=1-\eta_{\rm p}$ is the one in magnetic field (if one neglects magnetospheric radiation losses, which however can be a non negligible fraction of the total pulsar power especially in young $\gamma$-ray bright pulsars \citep{Iniguez-Pascual_Vigano+25a}). $Q_{\rm loss}(t)$ represents the total rate of particle losses (adiabatic, radiation, diffusion)  and $\mathcal{V}_{\rm pwn}(t)$ is the PWN volume. 
$P_{\rm pwn}^{\rm p}(t)$ and $P_{\rm pwn}^{\rm m}(t)$ are the PWN pressure of the plasma and magnetic component respectively (their sum is the total PWN pressure), while $W_{\rm pwn}(t)$ is the work done by the PWN as it expands into the ejecta.  In  Eq.~\ref{eq:e4}, $E_{\rm fil}(t)$ is the total energy of the filamentary network, including the thin shell bounding the PWN, while $E_{\rm ini}(t)$ is the energy associated to the ejecta that would have been contained inside the PWN volume. Losses due to radiative line emission from the filaments have been neglected. This is justified both by the fact that currently in the Crab nebula they are just a few percent of the work done by the PWN \citep{Davidson_Fesen85a,Pacini_Salvati73a}, and because, apart from possibly the very faint outer [OIII] skin, the emission of the filaments is well fitted by photo-ionization models, and not shock-heating  \citep{Sankrit_Hester97a,Henry_MacAlpine82A,MacAlpine_Satterfield08A,Charlebois_Drissen+10A,Owen_Barlow15A}, suggesting the ultimate source is the optical-UV synchrotron continuum and not the work of the PWN.  \\
\\
In the above equations we have assumed, as it is commonly done, that within the PWN there is no conversion of magnetic into particle energy. This is not generally true. It might be a good approximation for weakly magnetized systems, but as the magnetization rises, and magnetic energy approaches equipartition with the plasma dissipation of the former into the latter, associated to magnetic instabilities, is likely to be more efficient. This effect can be included adding coupling term in Eq.s~\ref{eq:e1}-\ref{eq:e2}. \\
\\
In order to close the system one needs to provide the following constitutive relations:
\begin{align}
   & E_{\rm pwn}^{\rm p}(t) = 3P_{\rm pwn}^{\rm p}(t)\mathcal{V}_{\rm pwn}(t)\,,\\
   & E_{\rm pwn}^{\rm m}(t) = 3P_{\rm pwn}^{\rm m}(t)\mathcal{V}_{\rm pwn}(t)\,,\label{eq:eosmag}\\
   & E_{\rm ini}(t)  = \frac{1}{2}\int_{\mathcal{V}_{\rm pwn}(t)} \rho_{\rm ej}(\boldsymbol{x},t)\left(\frac{r}{t}\right)^2 d^3x\,,
\end{align}
that amount to the assumptions that the magnetized plasma in the PWN  is relativistic, and 
that the ejecta are cold and in homologous expansion, with a density profile given by $\rho_{\rm ej}(\boldsymbol{x},t)$.  Note that in Eq.~\ref{eq:eosmag} we have assumed that the magnetic pressure acting at the boundary of the PWN is 1/3 of the average magnetic energy density. This was shown to be correct both for a fully turbulent and fully toroidal magnetic field in \citet{Bucciantini_Blondin+03a}. Moreover,  a closure for the mass and energy in the filamentary layer must be supplies. We opt for the following:
\begin{align}
    E_{\rm fil}(t) &= \frac{X}{2} M_{\rm fil}(t) \dot{R}_{\rm pwn}(t)^2 \,,\label{eq:efil}\\
    M_{\rm fil}(t) & = \int_{\mathcal{V}_{\rm pwn}(t)} \rho_{\rm ej}(\boldsymbol{x},t) d^3x\,,
\end{align}
where the parameter $X$ depends on the thickness, geometry and mass distribution in the filamentary layer, and ${R}_{\rm pwn}(t)$ represents the equatorial-radius of the PWN (for a spherical system its radius).
Finally one needs to assume a shape for the PWN volume. The simplest option is to assume a spherical nebula (but a prolate ellipsoid is no more difficult, if the prolateness is assumed to be time independent):
\begin{align}
    \mathcal{V}_{\rm pwn}(t) = \frac{4\pi}{3} R_{\rm pwn}(t)^3\,.
\end{align}
This completely closes the system. 
\section{Self-Similar Solution}
\label{sec:sefsim}
\subsection{Core-Envelope model for the SNR ejecta}
The simplest assumption one can make for the density structure in the homologous expanding ejecta is the so called core-envelope model \citep{Matzner_McKee99a,Truelove_McKee99a,Bandiera_Bucciantini+21a}:
\begin{align}
    \rho_{\rm ej}(r,t) = \begin{cases}
        \rho_{\rm c}(t) \left(\dfrac{r}{V_{\rm c}t} \right)^{-\delta} & {\rm for}\quad r <  V_{\rm c}t\,,\\
        \rho_{\rm c}(t) \left(\dfrac{r}{V_{\rm c}t} \right)^{-\omega} & {\rm for}\quad r \ge  V_{\rm c}t\,,
    \end{cases}
\end{align}
where, in terms of the SNR energy $E_{\rm sn}$ and total ejecta mass $M_{\rm ej}$:
\begin{align}
    V_{\rm c} &=\sqrt{\frac{2(5-\delta)(\omega - 5)}{(3-\delta)(\omega-3)}\frac{E_{\rm sn}}{M_{\rm ej}}}\,\label{eq:vcore},\\
    \rho_{\rm c} &= \frac{(3-\delta)(\omega-3)}{\omega-\delta} \frac{M_{\rm ej}}{4\pi}(V_{\rm c} t)^{-3}\,.
\end{align}
It can be shown that PWNe typically never reach the core-envelope transition at $V_{\rm c} t$ before reverberation begins, unless the PSR luminosity is very high (higher than that of the Crab PSR), or the supernova energy very small. As such, one can safely assume that the PWN will always expand in the core region of the ejecta. One can then use the so called core-rescaled quantities:
\begin{align}
    \tilde{M}_{\rm ej} &= 4\pi\int_0^{V_{\rm c}t}\rho_{\rm ej}(r,t)r^2 dr\,,\\
    \tilde{E}_{\rm sn} &= 2\pi\int_0^{V_{\rm c}t}\rho_{\rm ej}(r,t)\left(\frac{r}{t}\right)^2r^2 dr\,,
\end{align}
 related to the total mass and energy of the ejecta in a core-envelope model by:
\begin{align}
    \tilde{M}_{\rm ej} = \frac{\omega-3}{\omega-\delta}M_{\rm ej},\quad\quad\quad \tilde{E}_{\rm sn} = \frac{\omega-5}{\omega-\delta}E_{\rm sn}\,,
\end{align}
such that:
\begin{align}
    &V_{\rm c} =\sqrt{\frac{2(5-\delta)}{3-\delta}\frac{\tilde{E}_{\rm sn}}{\tilde{M}_{\rm ej}}}\,,\\
    &\rho_{\rm ej}(r<V_c t,t) = \frac{\left[(3-\delta) \tilde{M}_{\rm ej}\right]^{5/2}}{8\pi \sqrt{2}\,  t^3 \left[(5-\delta)\tilde{E}_{\rm sn}\right]^{3/2}}\left(\frac{r}{V_c t}\right)^{-\delta}.
\end{align}
If one neglects the radiative losses of the particles in the PWN, then the original equations simplify by introducing the total PWN pressure $P_{\rm pwn}$ and energy $E_{\rm pwn}$; if one does the further assumption that the PWN is spherical, the equations to solve become:
\begin{align}
   &P_{\rm pwn}(t) = \frac{E_{\rm pwn}(t)}{4\pi R(t)^3} = \frac{1}{4\pi R(t)^4}\int_0^t L_{\rm psr}(t')R(t')dt'\,,\label{eq:1st}\\
    &E_{\rm fil}(t) + E_{\rm pwn}(t) = E_{\rm ini}(t) + \int_0^t L_{\rm psr}(t')dt'\,, \label{eq:ess}
\end{align}
with:
\begin{align}
M_{\rm fil}(t) = \tilde{M}_{\rm ej}\left( \frac{R(t)}{V_{\rm c}t}\right)^{3-\delta}\quad {\rm and}\quad
E_{\rm ini}(t) = \tilde{E}_{\rm sn}\left( \frac{R(t)}{V_{\rm c}t}\right)^{5-\delta}.\label{eq:ss1}
\end{align}
If the PSR luminosity is constant, $L_{\rm psr}(t)=L_{\rm psr}=const$, it is easy to show that the above set of equations admits a self-similar solution in  the form:
\begin{align}
   R(t) = C t^{(6-\delta)/(5-\delta)}\,\label{eq:ss2},
\end{align}
with:
\begin{align}
    C^{5-\delta}\! =\!\!
\frac{2^{(5-\delta)/2} (5-\delta)^{(7-\delta)/2}
   (3-\delta)^{(\delta-3)/2} 
   (6-\delta)}{ (11-2 \delta) \left[X(6-\delta)^2-(3-\delta)(5-\delta)\right]}L_{\rm psr} \!\sqrt{\frac{\tilde{E}_{\rm sn}^{3-\delta}}{\tilde{M}_{\rm ej}^{5-\delta}}}.
   \label{eq:ceform}
\end{align}
One recovers the thin-shell limit by setting $X=1$ (see Appendix ~\ref{app:mform} for a comparison with the M-formalism).  

\subsection{Thick filamentary layer in the E-formalism}
Let us assume that the mass in the filamentary layer is distributed from an inner radius  $R_{\rm in}(t)$ to the nebular outer radius $R(t)$, according to: $\rho_{\rm fil}(r,t) \propto K(t)\; r^{-\zeta}$. The case $\zeta=0$ corresponds to a uniform distribution, while $\zeta=2$ corresponds to the assumption that the mass of the filaments is uniformly distributed in radius. The total  mass in the filaments will be:
\begin{align}
    M_{\rm fil}(t) &= 4\pi \int_{R_{\rm in}(t)}^{R(t)}\rho_{\rm fil}(r',t) r'^2 dr' = 4\pi K(t) \int_{R_{\rm in}(t)}^{R(t)} r'^{2-\zeta} dr'\nonumber\\
    &= \frac{4\pi}{3-\zeta} K(t)(R(t)^{3-\zeta}-R_{\rm in}(t)^{3-\zeta})\,.
\end{align}
In the Crab nebula there are good evidences that the filaments are in homologous expansion \citep{Martin_Milisavljevic+25a} with a velocity $\propto r$. So we assume that the velocity of the swept-up mass in the thick-layer is: $(r/R(t)) \dot{R}(t)$, equal to the expansion speed of the nebula $\dot{R}(t)$ at $R(t)$. One can define the total energy of the filamentary layer as:
\begin{align}
    E_{\rm fil}(t) &= 2\pi \int_{R_{\rm in}(t)}^{R(t)}\rho_{\rm fil}(r',t) \dot{R}(t)^2 \left(\frac{r'}{R(t)}\right)^2 r'^2 dr'\\
    &=\frac{3-\zeta}{2(5-\zeta)} M_{\rm fil}(t)\left(\frac{\dot{R}(t)}{R(t)}\right)^2\frac{R(t)^{5-\zeta}-R_{\rm in}(t)^{5-\zeta}}{R(t)^{3-\zeta}-R_{\rm in}(t)^{3-\zeta}}\,,
\end{align}
where the mass of the filamentary layer must be the same as the mass of the swept-up ejecta. By comparison with Eq.~\ref{eq:efil}, introducing $\kappa = R_{\rm in}(t)/R(t)$, one has:
\begin{align}
    X = \frac{3-\zeta}{5-\zeta}\; \frac{1-\kappa^{5-\zeta}}{1-\kappa^{3-\zeta}}\,.
\end{align}
In the case of a fully developed  ($R_{\rm in}=0$) and homogeneous ($\zeta=0$) filamentary layer, the  solution simplifies to:
\begin{align}
    E_{\rm fil}(t) = \frac{3}{10}M_{\rm fil}(t)\dot{R}^2(t)\,, \label{eq:thick2}
\end{align}
that is only $0.6$ times less than the case where all the swept up matter is in a thin shell. A thick/homologous distribution reduces kinetic energy at fixed $M_{\rm fil}$ and $\dot{R}$.\\
\\
One can show that for a fully developed homogeneous filamentary layer, choosing $\delta=0$, the radius of a thick-shell nebula is about 1.26 times larger than that of a thin-shell one, at the same age.
In \citet{Bucciantini_Amato+04a} it was shown that for a fully developed Rayleigh-Taylor filamentary network, penetrating inward up to about 0.7 times the radius of the PWN, the PWN outer radius was about 10\% larger that in the absence of instability. Using our formalism, and setting $R_{\rm in}(t) = 0.7 R(t)$, we find that the radius of a thick-shell nebula is about 1.1 times larger than the same radius of a thin-shell one, in full agreement with those numerical results.\\
\\
Using our approach it is possible to model systems where a shell still form but a sizable fraction $\xi_{\rm fil}$ of the swept-up mass ends in a thick homologous filamentary layer due to the shell fragmentation, by setting $X=1 - 0.4\xi_{
\rm fil}$.

   \begin{figure*}
        \centering
        \includegraphics[width=17cm]{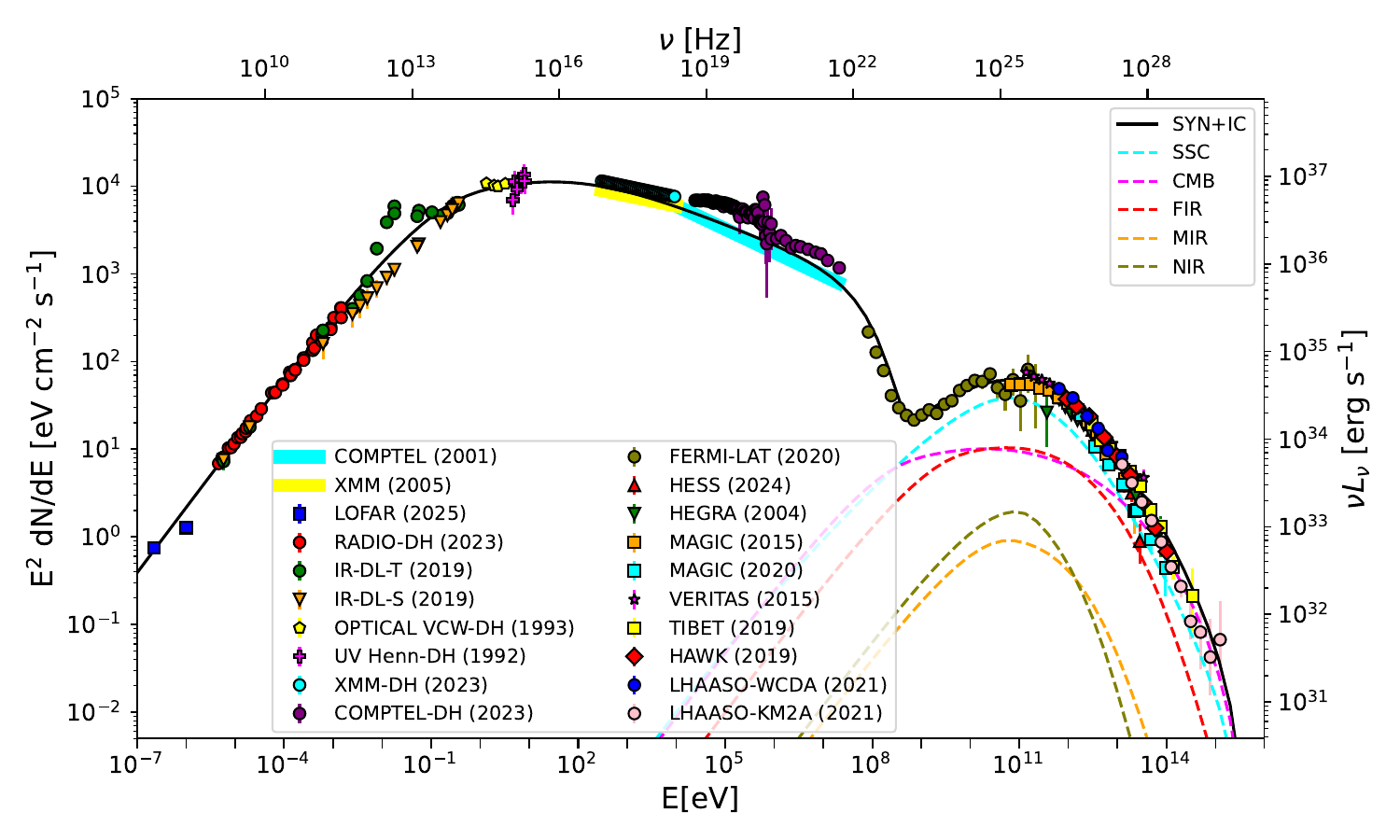}
        {\caption{Spectral energy distribution in the Crab nebula \citep{Aharonian_Ait+24a} and thick-shell model. The black solid line represents the expected total flux from the model of Tab~\ref{table:1}. Dashed lines are Inverse Compton on: self-synchrotron (SSC, cyan), CMB (magenta), FIR (red), MIR (orange), NIR (olive).   LOFAR \citep{Arias_Timmermann+25a}; RADIO-DH (compilation of radio data from \citet{Dirson_Horn23+}); IR-DL-T and IR-DL-S (total flux and estimated power-law synchrotron component from \citet{deLooze_Barlow+19a}); OPTICAL-VC-DH (original data from \citet{Veron-Cetty_Woltjer93a} with new extinction from \citet{Dirson_Horn23+}); UV-Henn-DH (original data from \citet{Hennessy_OConnel+92a} with new extinction); XMM \citep{Kirsch_Briel+05a}; COMPTEL \citep{Kuiper_Hermsen+01a}; XMM-DH and COMPTEL-DH (same data but rescaled to NUSTAR and INTEGRAL by \citet{Dirson_Horn23+}); FERMI-LAT \citep{Abdollahi+Acero+20a,Aharonian_Ait+24a}; HESS \citep{Aharonian_Ait+24a}; HEGRA \citep{Aharonian_Akhperjanian+04a}; MAGIC \citep{Aleksic_Ansoldi+15a,Magic+20a}; VERITAS \citep{Meagher_VERITAS51a}; TIBET \citep{Amenomori_Bao+19a}; HAWC \citep{Abeysekara_Albert+19a}; LHAASO \citep{Lhaaso+21a}.  }
         \label{fig:1}}
    \end{figure*}

\section{The Crab nebula}
\label{sec;crab}
The age of the Crab nebula is $971$~yr. However, to account for the fact that many of the structural data we are using for comparison are 10-25 year old, we adopt a fiducial age of $950$~yr.
Notice that 
the purpose of this section is to show how our model can account for both the broad emission properties of the PWN and the dynamical properties of the filamentary layer, and not to provide an optimal fit for this system. 

The size of the PWN has been measured from the edge of the radio and [OIII] emission
skin by \citet{Loll_Desch+13A}. Assuming an inclination on the plane of the sky of $25^\circ$, they fitted a prolate ellipsoid with a minor
axis of 1.43~pc and a major axis of 2.25~pc for a typical distance of 2 kpc. The radius of a sphere
with the same volume, which we take as the fiducial radius of our spherical model, is 1.66~pc. However, more recently \citet{Martin_Milisavljevic+21a}, using 3D tomography of
the filaments, find that the 3D shape differs from a simple ellipsoid, with clear indentations. They also find that most of the line emitting material is confined in a layer between 0.5 to 2~pc.
\citet{Bietenholz_Nugent15A} have measured the expansion speed and convergence dates for both the radio PWN ($1255\pm27$ CE) and the optical filaments ($1091\pm34$ CE). The more recent estimate for the mass of the filaments $7.2\pm0.5~M_\odot$ was given by \citet{Owen_Barlow15A}. For the PSR spin-down we adopt the standard values typically used in the literature \citep{Martin_Torres+12a,Bandiera_Bucciantini+23a} and based on the recent estimate by \citet{Lyne_Jordan+15a}: $L_{\rm psr}(t=0) = 3\times 10^{39}$~erg~s$^{-1}$ (for a standard moment of inertia $I_{\rm ns}=10^{45}$~g~cm$^{-2}$), with a spin-down time $\tau = 758$~yr and a braking index $n=2.51$. We further assume that the outer layer of the ejecta has $\omega=12$ \citep{Matzner_McKee99a}. Some preliminary studies on the possible range of masses and energetics of the SNR were recently done by \citet{Yang_Chevalier15a} in the non-radiative case, however the full implications of the low mass in the filaments were not properly addressed there and, as we are going to show, radiation losses cannot be neglected.
\\
\subsection{Adiabatic estimates}
If the PWN age $t_{\rm pwn}$ and its radius $R_{\rm pwn}$ are known, one can verify under what condition it is possible to assume that the expansion is still within  the core region of the ejecta. This sets for the Crab a lower limit on the supernova energy:
\begin{align}
     E_{\rm sn} &> \frac{3-\delta}{2(5-\delta)}\frac{\omega-3}{\omega-5}M_{\rm ej} \left(\frac{R_{\rm pwn}}{t_{\rm pwn}}\right)^2 \\
     &> 3.73 \frac{3-\delta}{(5-\delta)}\frac{M_{\rm ej}}{M_\odot}\times 10^{49}\;{\rm erg}\,.
\end{align}
Unless the supernova was tremendously sub-energetic, then it is fair to assume that the PWN is still inside the core region of the ejecta, even if quite close to its boundary. \\
\\
It is easy to show that in the limit of constant pulsar luminosity, the self similar solution, given by Eq.s-\ref{eq:ss1}-\ref{eq:ss2}-\ref{eq:ceform}, provides a relation for the swept-up mass that is just a function of the pulsar luminosity. One has:
\begin{align}
L_{\rm psr} \!=\!\! \frac{R_{\rm pwn}^{5-\delta}}{t_{\rm pwn}^{6-\delta}}\! \frac{ (11-2 \delta) \!\left[X(6-\delta)^2\!-\!(3-\delta)(5-\delta)\right]}{2^{(5-\delta)/2} (5-\delta)^{(7-\delta)/2}(3-\delta)^{(\delta-3)/2}(6-\delta)}
\!\!\sqrt{\frac{\tilde{M}_{\rm ej}^{5-\delta}}{\tilde{E}_{\rm sn}^{3-\delta}}}.
%
\end{align}
On the other hand the swept-up mass is:
\begin{align}
M_{\rm fil} &= \tilde{M}_{\rm ej}\left( \frac{R_{\rm pwn}}{V_{\rm c}t_{\rm pwn}}\right)^{3-\delta} = \frac{R_{\rm pwn}^{3-\delta}}{t_{\rm pwn}^{3-\delta}}\sqrt{\frac{\tilde{M}_{\rm ej}^{5-\delta}}{\tilde{E}_{\rm sn}^{3-\delta}}} \left( \frac{3-\delta}{2(5-\delta)}\right)^{(3-\delta)/2},
 \end{align}
 leading to:
\begin{align}
    M_{\rm fil} = \begin{cases}
        6.4\, M_\odot\;\dfrac{L_{\rm psr}}{3\times 10^{39}{\rm erg\,s^{-1}}}& {\rm for}\;\;\delta=0\;\,{\rm and}\;\,X=0.6,\\
        2.0 \,M_\odot \;\dfrac{L_{\rm psr}}{3\times 10^{39}{\rm erg\,s^{-1}}}& {\rm for}\;\;\delta=0\;\,{\rm and}\;\,X=1.0.
    \end{cases}
    \label{eq:masschange}
\end{align}
%
Note that for a given age and radius of a PWN, the amount of swept-up mass, i.e. the mass in the filaments, is just a function of the pulsar luminosity, and not of the supernova energy or ejecta mass, contrary to naive expectations. Given the accuracy in the timing of the Crab pulsar and the fact that the neutron star moment of inertia is supposed to vary in a narrow range \citep{Breu_Rezzolla16a}, this implies that there is no margin to adjust the mass in the filaments.
\\
\\
It is moreover evident that properly accounting for the formation of a filamentary layer is important in order to get a robust estimate of the swept-up mass, given that this can change by about a factor 3, as shown by Eq.~\ref{eq:masschange}. In any case however, the luminosities that are required are much higher than the average luminosity of the Crab pulsar, and closer to its initial one. For $\delta=1$ the masses are smaller by about 1/3. This also suggests, again contrary to naive expectations, that placing more mass at inner radii in the ejecta worsen the discrepancy between observations and theory.\\
\\
Assuming a fully developed filamentary network $X=0.6$, flat ejecta ($\delta =0$) and including the Crab spin-down, it is possible, in the non radiative case (solving the equations of Sect.~\ref{sec:eqns}), to reproduce both the PWN radius and the to get  $\sim 7~M_\odot$ in the filaments by assuming $E_{\rm sn} = 3.2\times 10^{50}$~erg, $M_{\rm ej}=12M_\odot$, and increasing the PSR spin-down power by a factor 1.57. These numbers are not totally unreasonable. Higher ejecta masses (higher progenitor masses) seems to be excluded by the composition of the Crab filaments \citep{Fesen_Rudie+08a, MacAlpine_Satterfield08A}, and ultimately as shown before, do not help much. The moment of inertia of the NS could be closer to $1.5\times 10^{45}$~g~cm$^2$, if the Crab pulsar was a more massive NS than expected from a low mass progenitor \citep{Sukhbold_Ertl+16a}, but still below $13~M_\odot$ \citep{Fesen_Shull+97A}.  Adopting a more realistic harmonic density model \citep{Matzner_McKee99a} makes things worse: to reproduce both the radius and the swept-up mass with no spin-down, one finds the combinations: $E_{\rm sn} = 3.85\times 10^{50}$~erg,  $M_{\rm ej}=12M_\odot$, $L_{\rm psr} = 4.7 \times 10^{39}$~erg~s$^{-1}$; or $E_{\rm sn} = 1.87\times 10^{50}$~erg,  $M_{\rm ej}=9M_\odot$, $L_{\rm psr} = 5.8 \times 10^{39}$~erg~s$^{-1}$. In the presence of spin-down, for $M_{\rm ej}=12M_\odot$, one needs to rise the PSR luminosity by a factor 2.33 over the standard value. \\
\\
\subsection{Radiative model}
The major problem arises when one tries to include radiation losses. In the presence of radiation losses, the energy of the PWN cannot be derived following Eq.~\ref{eq:1st}, but must be computed from the full energy distribution. We have for this purpose developed a high-order implicit upwind scheme (see App.s~\ref{app:evoene},\ref{app:kn} and \ref{app:averu} for a description of the method) that accounts for adiabatic, sychrotron and IC losses, both on the standard Galactic backgrounds \citep{Porter_Johannesson+17a} and of self-synchrotron in the full Klein-Nishina regime.  
It is easy to show that radiation losses integrated over the age of the nebula amount to more than half of the nebular energy, and that before $\sim 300$~yr half the injected power was lost to radiation. 
Neglecting radiation losses can then compromise substantially the overall dynamical estimates. 
Even adopting the more favorable core-envelope profile for the ejecta, and assuming a standard broken power-law injection (with high and low energy indexes $p_{\rm h} = 2.35$ and $p_{\rm l}=1.5$ respectively, and break at $\gamma_{\rm br} = 5\times 10^5$,  \citealt{Bandiera_Bucciantini+23a}), it
is not possible to reproduce the net synchrotron luminosity ($\sim 1\times 10^{38}$~erg~s$^{-1}$) and the measured swept-up mass, with $M_{\rm ej}\le 12M_\odot$, unless the pulsar spin-down power is augmented by a factor $\simeq 3.3$ (to increase the energetics of the PWN), and simultaneously the magnetic field in the nebula is lowered to values $B \sim 30~\mu$G (to reduce the losses).
%
But  in this case, 
one cannot then reproduce the observed spectrum of the Crab nebula: in particular the computed IC component largely overestimates the measured value. 

It is obvious that there does not seem any way of reproducing both the current spectrum of the Crab nebula, as well as its size and the mass of the filamentary network, once radiation losses are included. Even if one assumes that magnetization was much higher in the past, in order to prevent radiation losses in the very early stages to dominate the energetics (if most of the energy in the early nebula was in the form of magnetic field, then radiation losses, which only affect particles, would not be so detrimental), it is still not possible to reproduce the spectrum, because initial magnetization must be pushed up so much ($\eta_{\rm p} = 0.01$), that one ends with a magnetic field $\sim 400~\mu$G at present, and again is unable to reproduce the spectrum. \\
\\
In the standard thin-shell approach (solving the equations presented in Sect.~\ref{sec:eqns} with $X=1$) assuming $E_{\rm sn}=10^{51}$~erg, $M_{\rm ej}=8.0~M_\odot$ and adopting  a magnetic fraction $\eta_{\rm b}=0.02$, one can get the correct radius of 1.66~pc at 950~yr, with a magnetic field $\sim 100~\mu$G, but the mass in the filaments is only $0.45~M_\odot$. If one assumes a fully developed filamentary layer ($X=0.6$) then one can reproduce the nebula using the same values as before but just lowering the supernova energy to $E_{\rm sn}=0.47\times10^{51}$~erg. This lowers the values of $V_{\rm c}$ according to Eq.~\ref{eq:vcore}, and given that, for $\delta=0$, $M_{\rm fil}\propto E_{\rm sn}^{-3/2}$, this leads to a mass in the filaments of $\sim 1.4~M_\odot$.\\
\\
Unfortunately, it is evident that even allowing for a fully developed filamentary layer, the coupling between the PWN and the ejecta is too strong, and one cannot recover the correct filamentary mass, unless one changes substantially the PSR spin-down power, with major consequences on the spectral fitting. It seems unlikely that the PSR spin-down power was much different than what we expect at earlier time. The Crab PSR has been timed constantly for 50 years and shows one of the cleanest spin-down history among PSRs; Moreover, since its discovery in the 18th century, there has been no evidence for any major variation in the luminosity of the nebula.
\subsection{Decoupling the ejecta from the PWN}
A possible way out is to lower the coupling between the swept-up ejecta and the PWN.  In writing Eq.~\ref{eq:thick2}, we made the assumption that the swept-up ejecta were in homologous expansion with an outer speed equal to the expansion speed of the PWN. However, if coupling is inefficient, their outer expansion speed might be smaller than the expansion speed of the PWN (any value between $V_{\rm pwn}$ and $V_{\rm c} R_{\rm pwn} / R_{\rm c}$ is in principle allowed). The idea is that the PWN penetrates the ejecta partially dragging them instead of pushing them.  We can introduce a dragging coefficient $\xi_{\rm drag}$ such that  for a fully developed filamentary layer one has:
\begin{align}
    \dot{R}^2 = \xi_{\rm drag}^2\frac{10}{3}\frac{E_{\rm fil}(t)}{M_{\rm fil}(t)}\,.
\end{align}
This is consistent with the fact that the filaments are seen to expand slower (have a later convergence time) than the radio PWN. One can get $7~M_\odot$ in the filaments by setting $\xi_{\rm drag} =1.141$, with $E_{\rm sn}=0.3\times10^{51}$~erg and $M_{\rm ej}=12~M_\odot$. This amount to a relative difference of $\sim15$\% in the convergence time  between the radio nebula and optical filaments.
In practice, the swept-up ejecta have not really been "swept-up" in any meaningful sense, and the non-thermal PWN has simply break through them.  

Note that what really counts in the evolution of the PWN are not the value of $X$ and $\xi_{\rm drag}$ separately, but the combination $X/\xi_{\rm drag}^2$. For example the same result of the fully developed filamentary layer with $\xi_{\rm drag} =1.141$, can be obtained using a layer extended over half of the PWN, $X=0.664$, with  $\xi_{\rm drag} =1.200$. This also suggests that it might be possible to modify the equations of the standard M-formalism by introducing an effective inertial mass for the thin-shell in the momentum conservation law, that is smaller than the one derived from the mass conservation law. In this respect the E-formalism can provide a physical interpretation, in terms of decoupling, or thickness of such difference.  However, one should not naively think to use the same value of $X$ obtained in the E-formalism. Modifying the momentum conservation law of the M-formalism (e.g. Eq.~2 in \citet{Bandiera_Bucciantini+20a}) with an effective inertia according to:
\begin{align}
    \frac{d}{dt}&[X_{\rm val}M(t) \dot{R}(t)]=\nonumber\\
    &=4\pi R(t)^2\left[P_{\rm pwn}(t) - \rho_{\rm ej}(r,t)\left(\dot{R}(t)-\frac{R(t)}{t} \right)\frac{R(t)}{t}\right],\label{eq:mmod1}
\end{align}
in the limit of constant pulsar spin-down luminosity and assuming that $\rho_{\rm ej}(r,t) \propto r^{-\delta}$, one can find a relation between the value of $X$ in the E-formalism and $X_{\rm val}$:
\begin{align}
    X_{\rm val} = \frac{2 \delta ^3-\delta ^2 (5 X+23)+\delta  (60 X+86)-15 (12 X+7)}{2 (\delta -6)^2 (\delta -4)}.\label{eq:mmod2}
\end{align}
\\
We have verified that over the typical range of parameters that characterize a PWN-SNR system \citep[e.g.][]
{Bandiera_Bucciantini+23a}, as long as radiation losses are neglected, the modification to the M-formalism provided by Eq.s~\ref{eq:mmod1}-\ref{eq:mmod2}, give
a  radial evolution that differs at most by 10\% (for $X=0.6$, and for very short spin-down times) from the one computed in the correct E-formalism, up to the beginning of the reverberation. 
\\
In Fig.~\ref{fig:1} we show the spectrum computed with our model, as defined in Tab.~\ref{table:1}, computed using the \texttt{naima}\footnote{\url{https://naima.readthedocs.io/en/latest/}} package \citep{Zabalza15a}. The model assumes a standard broken-power-law for the particle injection. The purpose of this is not to present an optimal fit of the Crab (structural and observational data in the Crab nebula are typically taken at different times, and a proper fit done using an evolutionary model, should account for this, something that is typically not done) but to show that a model based on our formalism can perform as satisfactory as any of the existing 1ZTS approaches. The main discrepancies between the data and our model are mostly in two bands. In the $\mu$m-mm band,  our model suggests the synchrotron component should display a non negligible level of curvature (in part due to the presence of a spectral break in the IR, in part due to evolutionary effects), for which there is some evidence in recent James-Webb Telescope observations \citep{Temim_Laming+24a}, but which was not considered in the derivation by \citet{deLooze_Barlow+19a}. Above 100~TeV our model, if matched to the more recent X-ray data by \citet{Dirson_Horn23+}, tends to overpredic the LHAASO fluxes \citep{Lhaaso+21a}. This comes from the fact that the IC above 100~TeV is mostly due to CMB up-scattered by particles responsible for synchrotron emission in the hard X-ray and MeV range. These particles are typically located in the inner region of the PWN, close to the termination shock, and as such they tend to be embedded in a local magnetic field higher than the average one. 
One zone models, that are forced to use an  average magnetic filed, tend to over-estimate the number of high-energy particles, and so their contribution to the IC. It is possible to partly balance this using a very soft high-energy injection (our model requires $p_{\rm h} =2.43$, while optical observation seems to suggest values closer to $2.2-2.3$), but one need to be careful not overshooting the optical synchrotron flux. Recent estimates of the extinction in optical \citep{Dirson_Horn23+} have lowered the PWN luminosity by about 30\%, 
making this constraint a little more binding than before.\\
\\
It is interesting to note that the ratio $R_{\rm pwn} /V_{\rm c} t_{\rm pwn}$ is $\sim 93\%$, meaning that the PWN has reached the edge of the core region in the ejecta. This is consistent with the lack of any extended flat density halo surrounding the Crab nebula, and with other indications that the system is expanding inside ejecta with a steeper radial profile \citep{Sankrit_Hester97a,Tziamtzis_Schirmer+09a,Lundqvist_Tziamtzis12a}.

\begin{table}[h!]
\caption{Input and output parameters for the thick-shell radiative model for the Crab nebula shown in Fig.~\ref{fig:1}}                 
\label{table:1}    
\centering                        
\begin{tabular}{l c}      
\hline\hline               
Input Parameter & {} \\         
\hline     \\                
   Age & 950~yr \\
   $L_{\rm psr}(t=0)$ & $3\times 10^{39}$~erg~s$^{-1}$ \\    
   $\tau_{\rm sd}$ & 758~yr \\
   $n$ & 2.51 \\
   $\eta_{\rm b}$ & 0.024\\
   $M_{\rm ej}$ & 12~$M_\odot$ \\
   $E_{\rm sn}$ & $3\times 10^{50}$~erg \\
   $\delta$ & 0 \\
   $\omega$ & 12 \\
   $X$ & 0.6 \\
   $\xi_{\rm drag}$ & 1.141 \\
   $\gamma_{\rm min}$ & $5$ \\
   $\gamma_{\rm max}$ & $8\times 10^9$\\
   $\gamma_{\rm br}$ &  $6\times 10^5$\\
   $p_{\rm h}$ &  1.5\\
   $p_{\rm l}$ &  2.43\\
   $T(FIR)$ & 31~K\\
   $U(FIR)$ & 0.322~eV~cm$^{-3}$\\
   $T(MIR)$ & 350~K\\
   $U(MIR)$ & 0.051~eV~cm$^{-3}$\\
   $T(NIR)$ & 3650~K\\
   $U(NIR)$ & 0.441~eV~cm$^{-3}$\\
   \\
\hline  
Output Parameter & {} \\         
\hline    \\          
   $R_{\rm pwn}(950~{\rm yr})$ & 1.66~pc\\
   $M_{\rm fil}(950~{\rm yr})$ & 7.02~$M_\odot$\\
   Conv. Date (Radio) & 1210 CE\\
   Conv. Date (Filaments) & 1090 CE\\
   $B(950~{\rm yr})$ & 107~$\mu$G\\
\end{tabular}
\end{table}

\section{Conclusions}
\label{sec:conclusions}
We have shown that it is possible to formulate the problem of the evolution of a wind bubble as a one zone object, using energy conservation instead of momentum conservation at the outer boundary. This new energy based formalism allows us to include effects like the development of a thick layer of mixed filaments, asphericity or partial dragging. In particular we were able to reproduce the structural properties of the Crab nebula, like the large mass of the filaments, and the differences in convergence age between the radio and the filaments themselves, together with its spectrum. \\
\\
We summarize here the main results:
\begin{itemize}
    \item It is not possible with the standard 1ZTS model to reconcile the energetics of the PSR, the inferred mass of the filaments, and the observed spectrum of the Crab nebula.
    \item Even allowing for a thick filamentary layer, one must reduce the coupling between the PWN and the filaments to very small values, in order to fit the inferred masses.
    \item The E-formalism can easily be extended to include more components (like ions in the PSR winds) or other kind of losses (e.g. diffusion) that might be relevant for very high energy particles, as well as other geometrical properties like non-uniform distribution of masses.
    \item We show how the effect of partial coupling, or the development of a thick layer, can be included in the standard 1ZTS models by the introduction of a reduced inertia for the swept-up shell.
    \item Our model of the Crab nebula is in agreement with the lack of any flat density halo extending outside the nebula.
    \item In the $\mu$m-mm energy range the sychrotron spectrum has curvature, and simple power-law models tend to underestimate its contribution.
    \item Emission above 100~TeV in one zone models tends to be overestimated because of the use of an average magnetic field to fit emission in the hard X-ray and MeV range.  
    \item Care has to be taken in evolving the particle distribution in order to ensure energy conservation, and in this respect standard 1st-order schemes prove quite inefficient.
    \item Computation of self-synchrotron-Compton losses can be formulated in a fully algebraic-vectorizable way even in the full Klein-Nishina regime, by defining an effective energy density of the photon-field. 
\end{itemize}
The fact that under complete-coupling it is not possible to reproduce the observed mass of the filamentary layer, and that partial dragging must be included, raises the question on the origin of such filaments: if the swept-up ejecta are not swept-up, and fingers do not come from Rayleigh-Taylor instability, then a different picture has to be invoked, questioning the canonical interpretation. 
A possibility is to assume that the density in the core region of the ejecta, instead of being a smooth power-law, is more stratified, with an inner low density region, a cavity, and an outer high density region with pre-existing clumps, that the PWN will leave mostly behind as it expands. This agrees with claims about RT fingers being suppressed or strongly altered by the nebular magnetic field \citep{Bucciantini_Amato+04a,Stone_gardiner07a}. Another possibility is to invoke some more complex circumstellar environment \citep{Smith13A,Meyer_Meliani+24a,Meyer_Torres25a}. \\
\\
In the light of the present discussion, a related question is the reliability of standard 1ZTS models when used to perform PWN-SNR population studies, or to infer properties of the progenitor star. 
The strength of classical 1ZTS models (or any 1D model) is that they allow to relate the PWN evolution to the parent SNR properties in a very simple way (the only  parameters that matters are $E_{\rm sn}$, $M_{\rm ej}$ and the ISM density). 
As of today 1ZTS models, or even those that use a less approximated evolution of the PWN and SNR shell, as suggested recently in \citet{Bandiera_Bucciantini+23b}, remain the only way to model the spectral evolution of middle-aged and old system post reverberation. How one should factor in the presence of an extended layer of mixed material is not clear, and deserve further investigation. In particular, our findings suggest that both the assumption of complete-coupling as well as the thin-shell one, might introduce substantial biases to the point that is not clear how robust is the relation between the parameters $M_{\rm ej}$ or $E_{\rm sn}$ with the actual mass of the true ejecta, or energy of the parent supernova. How this might impact many of recent population models \citep[e.g.][]{Fiori_Olmi+22a}, and if its is or not relevant, rest to be understood.  

\begin{acknowledgements}
The authors thanks D.~F. Torres for fruitful comments and suggestions on this work. The authors acknowledge partial support by the INAF Mini-grant HYPNOTIC87A: "Hidden Young Pulsar Nebula Occupying The Inner Core of 87A" and by the European Union – NextGenerationEU RRF M4C2 1.1 grant PRIN-MUR 2022TJW4EJ.
\end{acknowledgements}

\bibliographystyle{aa}
\bibliography{article_1}
\begin{appendix}
\section{Comparison with the M-formalism}
\label{app:mform}
It is not difficult to show that, in the limit of a constant PSR spin-down luminosity, the M-formalism \citep{Bandiera_Bucciantini+20a} admits a self-similar solution analogous to the one of the E-formalism \citep{Reynolds_Chevalier84a}:
\begin{align}
    R(t) = C t^{(6-\delta)(5-\delta)}\,,
\end{align}
with:
\begin{align}
        C^{5-\delta} =\frac{2^{(3-\delta)/2}\,
      (5-\delta)^{(9-\delta)/2}
     }{(3-\delta)^{(3-\delta)/2}\,(11-2 \delta) (9-2\delta)}\,L_{\rm psr}\sqrt{\frac{\tilde{E}_{\rm sn}^{3-\delta}}{\tilde{M}_{\rm ej}^{5-\delta}}}.
\end{align}
This, however, is different form the one of Eq.~\ref{eq:ceform} in the equivalent thin-shell case $(X=1)$.
The origin of the discrepancy, is due to the fact that the energy in the filaments (or shell) $E_{\rm fil}(t)$ will have both a kinetic and an internal-energy term, while Eq.~\ref{eq:efil} only accounts for the kinetic component. In the thin-shell limit the thermal component cannot be determined and it is typically neglected. However, it is easy to show that this is not correct. From a physical point of view the assumption that the shell is infinitely thin, is equivalent to the assumption that the ejecta have infinite compressibility, or, stated in other terms, that their adiabatic coefficient  $\Gamma\rightarrow 1$. It is not difficult to show \citep{Jun98a} that the shell thickness scales as $R(t)(\Gamma-1)$, while its pressure is $\simeq P_{\rm pwn}(t)$, almost constant across the shell, and its internal energy density then scales as $\simeq P_{\rm pwn}(t) / (\Gamma-1)$, such that the integrated internal energy converges to a finite value, scaling as $P_{\rm pwn}(t) R(t)^3$. \\
\\
One can account for this by replacing, in the thin-shell case $X=1$, Eq.~\ref{eq:ceform} with:
\begin{align}
    E_{\rm fil}(t) =\frac{M_{\rm fil(t)}}{2} \dot{R}(t)^2 +  AE_{\rm pwn}(t)\,,
\end{align}
where the parameter $A$ can be found by requiring the solution of the E-formalism:
\begin{align}
    C^{5-\delta} &=
    \frac{2^{\frac{5}{2}-\frac{\delta}{2}} (\delta-5)^{\frac{7}{2}-\frac{\delta}{2}}
   (\delta-3)^{\frac{\delta-3}{2}} (
   (6-\delta)-A(5-\delta))}{ (2 \delta-11) (4 \delta-21)}\times\nonumber\\
   &\quad\quad\times  L_{\rm psr}\sqrt{\frac{\tilde{E}_{\rm sn}^{3-\delta}}{\tilde{M}_{\rm ej}^{5-\delta}}}\,,
\end{align}
to match the one of the M-formalism. The matching condition is:
\begin{align}
      A=  \frac{3-\delta}{2 \left(2 \delta^2-19 \delta+45\right)}\,,
\end{align}
    with $A$ confined  between $A=1/30$ at $\delta =0$ and $\delta =2$ and $A\simeq 1/27.9$ at $\delta =1.25$ ($A= 1/28$ at $\delta =1$).
 \noindent
In general setting $A=0$ leads to errors in the value of the PWN radius that are always $<1$\%. While using a fixed value $A=1/30$ gives an error that, at most, is $0.5$\% at $\delta=1$. Now it has been shown that for self-similar expansion of a piston (the PWN) inside ejecta themselves in homologous expansion, the hydrodynamical solution is self-similar \citep{Jun98a}. Then one can compute (as a function of the adiabatic coefficient) the structure of the swept-up shell.  Once the structure is known, it is easy to compute its internal energy. By taking solutions in the limit $\Gamma \rightarrow 1$, one can show that the total internal energy converges to the values derived by matching the solutions in the  E- and M-formalism. However, in general, neglecting the thermal contribution leads to negligible deviations in the global radial evolution.

\section{Prolate  nebulae}
\label{app:prolate}
One of the main advantages of the  E-formalism is that it can easily be extended to account for a non-spherical PWN shape. Here we will briefly illustrate how ellipticity can be included. To a good approximation the shape of a young PWN might be described by a prolate-ellipsoid:
\begin{align}
    R(t,\theta)^2 = \frac{R_{\rm eq}^2(t)}{1-e^2\cos^2{\theta}}\,, \quad {\rm with} \quad e^2 = \frac{R_{\rm pol}^2(t)-R_{\rm eq}^2(t)}{R_{\rm pol}^2(t)}\,,
\end{align}
where the labels $_{\rm eq}$ and $_{\rm pol}$ refer to the the equatorial and polar radii respectively. If one assumes that the ellipticity $e$ remains constant in time, one can easily make use of the following integrals:
\begin{align}
    \mathcal{V}^n &= 2\pi \int_0^\pi \!\!\!\int _0^{R(\theta)} \!\!\! r^n r^2 \sin{\theta} {\rm d}r {\rm d}\theta \nonumber\\
    &= \frac{4\pi}{3+n} {\rm {}_2F_1}\left[\frac{1}{2},\frac{3+n}{2},\frac{3}{2},e^2\right]R_{\rm eq}^{3+n}\,\label{eq:b2},
\end{align}
where ${\rm {}_2F_1}$ is the Hypergeometric function (it can be expressed in terms of simpler functions for $n$ multiple of 2, and for all integer $n\le 1$). The mass in the filaments, is given by the integral over the PWN volume of the density of the ejecta, and it will be:
\begin{align}
    M_{\rm fil}(t) = 2\pi \int_0^\pi \int_{R_{\rm in}(t,\theta)}^{R(t,\theta)}\rho_{\rm fil}(t,\theta,r') r'^2 \sin{\theta} d\theta dr' \label{eq:b3}\,.
\end{align}
Now one can make some simplifying assumptions for the distribution of the density of the filamentary layer.  First we assume that the thick-layer is self similar $R_{\rm in}(\theta,t)= \xi R(\theta,t)$, and that the density in the layer is uniform in both radius and angle, $\rho_{\rm fil}(t,\theta,r') = K(t)$. Then from Eq.~\ref{eq:b2} and Eq.~\ref{eq:b3} one has:
\begin{align}
    M_{\rm fil}(t) = \frac{4\pi}{3}{\rm {}_2F_1}\left[\frac{1}{2},\frac{3}{2},\frac{3}{2},e^2\right] K(t)R_{\rm eq}(t)^{3}(1-\xi^3)\,.
\end{align}
Analogously the energy in the filamentary layer, assuming homologous expansion, and no relative dragging ($\xi_{\rm drag}=1$), will be given by:
\begin{align}
    E_{\rm fil}(t)\! =\! \pi \!\!\int_0^\pi \!\!\!\int_{R_{\rm in}(t,\theta)}^{R(t,\theta)} \!\!\!\!\!\!\!\!\!\!\!\rho_{\rm fil}(t,\theta,r') r'^2 \dot{R}^2(\theta,t)\left(\frac{r'}{R(\theta,t)} \right)^2 \sin{\theta} d\theta dr',
\end{align}
and, under the same assumptions done for the density, one gets:
\begin{align}
    E_{\rm fil}(t)\! = \! \frac{2\pi}{5}{\rm {}_2F_1}\left[\frac{1}{2},\frac{5}{2},\frac{3}{2},e^2\right] K(t)\left(\frac{\dot{R}_{\rm eq}(t)}{R_{\rm eq}(t)}\right)^2 \!\!R_{\rm eq}(t)^{5}(1-\xi^5).
\end{align}
For a fully developed filamentary layer ($\xi=0$):
\begin{align}
    E_{\rm fil}(t) = \frac{3}{10}\frac{3-2e^2}{3-3e^3}M_{\rm fil}(t)\dot{R}^2_{\rm eq}(t)\,,
\end{align}
Now the mass of the filamentary layer is equal to the swept up mass of the ejecta, and as long as the PWN is confined to the ejecta core, Eq.~\ref{eq:b2} and Eq.~\ref{eq:ss1} give:
\begin{align}
\rho_{\rm c}(t) R_{\rm c}(t)^\delta \frac{4\pi}{3-\delta} {\rm{}_2F_1}\left[\frac{1}{2},\frac{3-\delta}{2},\frac{3}{2},e^2\right] R_{\rm eq}(t)^{3-\delta}
\end{align}
and recalling that:
\begin{align}
\tilde{M_{\rm ej}}=\frac{4\pi}{3-\delta} \rho_{\rm c}(t)R_{\rm c}(t)^3\,, \quad{\rm with}\quad R_{\rm c}(t) = V_{\rm c} t\,,
\end{align}
one finds for the swept-up mass:
\begin{align}
\tilde{M_{\rm ej}} \left(\frac{R_{\rm eq}(t)}{V_{\rm c} t}\right)^{3-\delta}  {\rm {}_2F_1}\left[\frac{1}{2},\frac{3-\delta}{2},\frac{3}{2},e^2\right]\,, 
\end{align}
and for the ejecta energy:
\begin{align}
E_{\rm ini}(t) = \tilde{E_{\rm sn}} \left(\frac{R_{\rm eq}(t)}{V_{\rm c} t}\right)^{5-\delta}  {\rm {}_2F_1}\left[\frac{1}{2},\frac{5-\delta}{2},\frac{3}{2},e^2\right]\,. 
\end{align}
Substituting these results into the general equations of Sect.~\ref{sec:eqns},   one can show that the equations for the equatorial radius reduce to:
\begin{align}
    \frac{{\rm d} E_{\rm pwn}(t)}{{\rm d} t} &= L_{\rm psr}(t)-\frac{4\pi}{3}\frac{3P_{\rm pwn}(t)R_{\rm eq}^2(t)}{\sqrt{1-e^2}}\frac{{\rm d} (R_{\rm eq}(t))}{{\rm d}t} \nonumber\\
    &= L_{\rm psr}(t) -\frac{E_{\rm pwn}(t)}{R_{\rm eq}(t)} \frac{{\rm d} (R_{\rm eq}(t))}{{\rm d}t} \,,\\
    \frac{{\rm d} W(t)}{{\rm d} t} &= \frac{E_{\rm pwn}(t)}{R_{\rm eq}(t)} \frac{{\rm d} (R_{\rm eq}(t))}{{\rm d}t}\,.
\end{align}
In the limit of constant PSR luminosity, and following the same strategy of Sect.~\ref{sec:sefsim}, one can show that the evolution of the equatorial radius  remains self similar, with the same power-law dependence of the spherical case.
Now we can introduce the following quantity:
\begin{align}
 Y[\delta,e] =  \frac{3}{5}\frac{{\rm{}_2F_1}\left[\frac{1}{2},\frac{5}{2},\frac{3}{2},e^2\right]}{{\rm {}_2F_1}\left[\frac{1}{2},\frac{3}{2},\frac{3}{2},e^2\right]}  \frac{{\rm{}_2F_1}\left[\frac{1}{2},\frac{3-\delta}{2},\frac{3}{2},e^2\right]}{{\rm {}_2F_1}\left[\frac{1}{2},\frac{5-\delta}{2},\frac{3}{2},e^2\right]} \,,
\end{align}   
which can easily be generalized to $\xi\neq 1$, or for filamentary layers with a non uniform radial density distribution.
Then the evolution will be given by:
\begin{align}
    &\frac{Y[\delta,e]\tilde{M}_{\rm ej}}{2}\left(\frac{R_{\rm eq}(t)}{V_{\rm c} t}\right)^{3-\delta}\!\!\!\!\!\!\dot{R}_{\rm eq}^2(t)  {\rm {}_2F_1}\left[\frac{1}{2},\frac{5-\delta}{2},\frac{3}{2},e^2\right]  =  \nonumber\\
    &=\quad W(t)+{\rm {}_2F_1}\left[\frac{1}{2},\frac{5-\delta}{2},\frac{3}{2},e^2\right] \tilde{E}_{\rm sn} \left(\frac{R_{\rm eq}(t)}{V_{\rm c} t}\right)^{5-\delta}\,.
\end{align}
Assuming that $R_{\rm eq}(t)=C t^{(6-\delta)/(5-\delta)}$  the solution becomes:
\begin{align}
    C^{5-\delta} &=
    \frac{2^{\frac{5}{2}-\frac{\delta}{2}} (\delta-5)^{\frac{7}{2}-\frac{\delta}{2}}
   (\delta-3)^{\frac{\delta-3}{2}} 
   (6-\delta)}{ (2 \delta-11) (Y[\delta,e](6-\delta)^2-(3-\delta)(5-\delta))} \times\nonumber\\
   &\times \frac{1}{{\rm {}_2F_1}\left[\frac{1}{2},\frac{5-\delta}{2},\frac{3}{2},e^2\right]}L_{\rm psr}\sqrt{\frac{\tilde{E}_{\rm sn}^{3-\delta}}{\tilde{M}_{\rm ej}^{5-\delta}}}\,.
\end{align}
$Y[\delta,e]$ is an increasing function of $\delta$ and $e$, rising from $3/5$ for $\delta =0$ independent of $e$, and for $e=0$ independent of $\delta$, up to $0.66$ for $\delta =2$ and $e=0.8$. While ${}_2F_{1}\left[1/2,(5-\delta)/2,3/2,e^2\right]$ goes from 0 at $e=0$, to 2.65 at $e=0.8$ and $\delta =0$ (1.65 at $e=0.8$ and $\delta =2$). This corresponds to a change by at most a factor 1.2 in the equatorial radius with respect toa spherical case.
\section{Second-order upwind implicit solver for the particle evolution}
\label{app:evoene}
In the presence of radiative losses, the total particle energy content of the PWN is not computed directly from Eq.~\ref{eq:e1}, but it is derived from the particle distribution function $f(\gamma)$. The latter, in the one zone model (assuming isotropy and homogeneity), evolves according to the continuity equation in energy-space:
\begin{align}
\frac{ \partial f(\gamma)}{\partial t} + \frac{ \partial \dot{\gamma}f(\gamma)}{\partial \gamma} = J(\gamma)\,,
\end{align}
where $J(\gamma)$ is the injection term, and $\dot{\gamma}$ describe the net effect of losses, adiabatic plus radiative \citep{Gelfand_Slane+09a,Bucciantini_Arons+11a,Martin_Torres+12a}. Due to the fact that the timescale for radiative losses can differ by many orders of magnitude with respect to the PWN evolutionary timescale, this equation is typically discretized according to the following implicit scheme:
\begin{align}
\frac{f^n_i -f^{n-1}_i}{\Delta t} + \frac{ \dot{\gamma}_{i+\frac{1}{2}}f^n_{i+\frac{1}{2}}-\dot{\gamma}_{i-\frac{1}{2}}f^n_{i-\frac{1}{2}}}{(\gamma_{i+\frac{1}{2}} -\gamma_{i-\frac{1}{2}})} = J_i^n\,,
\end{align}
where $n$ is the time step, while $i$ is the index of the energy bin. In this formulation the discretized equation ensures particle conservation to machine accuracy. However, it does ensure energy conservation only to interpolation accuracy. This means that, in the absences of radiation losses, the total energy computed from the particle distribution function does not match the one computed according to Eq.~\ref{eq:1st}.\\
\\
Now a first-order upwind definition of the $\gamma$-interface fluxes is:
\begin{align}
    \dot{\gamma}_{i+\frac{1}{2}}f^n_{i+\frac{1}{2}} =
    \begin{cases}
        \dot{\gamma}_{i+\frac{1}{2}}f^n_i \quad{\rm for}\quad \dot{\gamma}_{i+\frac{1}{2}} \ge 0 \,,\\
        \dot{\gamma}_{i+\frac{1}{2}}f^n_{i+1} \quad{\rm for}\quad \dot{\gamma}_{i+\frac{1}{2}} \le 0\,,
    \end{cases}
\end{align}
and the implicit equation reduces to the solution of a tri-diagonal matrix ${\rm A}$ with:
\begin{align}
    {\rm main} &= {\rm A}_{i,i} = \left[1+\frac{\Delta t}{\Delta\gamma_i} \left({\rm Max}[\dot{\gamma}_{i+\frac{1}{2}},0] - {\rm Min}[\dot{\gamma}_{i-\frac{1}{2}}, 0]  \right) \right] ,\\
    {\rm upper} &= {\rm A}_{i,i+1} = \frac{\Delta t}{\Delta\gamma_i} \left({\rm Min}[\dot{\gamma}_{i+\frac{1}{2}}, 0] \right) ,\\
    {\rm lower} &= {\rm A}_{i,i-1} = -\frac{\Delta t}{\Delta\gamma_i} \left( {\rm Max}[\dot{\gamma}_{i-\frac{1}{2}},0]\right), 
\end{align}
with the following boundary conditions  $f^n_{{\rm max}+1} =0$ and $f^n_{1/2} = f^n_{1} (\gamma_{1/2}/\gamma_1)^p$, where $p$ is the low-energy power-law index, and $\gamma_{1/2} = \gamma_1-\Delta\gamma_1/2$. This tri-diagonal can easily be inverted using a fast Thomas algorithm \citep{Press_Flannery+86a}.\\
\\
 However, depending of the injection spectrum, for a typical first order scheme with 150 logarithmically spaced energy bins, in the range $\gamma=10-10^{11}$, the discrepancy on the energy can be as high as 15\%. To get an error smaller than 1\% one needs to increase the number of the energy bins by a tenfold. This becomes prohibitive (especially when one wishes to include self-synchrotron-Compton) because implicit schemes are typically $\mathcal{O}(N^2)$ with respect to the number of bins.\\
 \\
One can use a second-order interpolation to increase accuracy. To do this, we compute the local power-law slopes of the distribution as:
\begin{align}
    p^{\rm r}_i = \log{(f_{i+1}/f_{i})} / \log{(\gamma_{i+1}/\gamma_{i})}\,\\
    p^{\rm l}_i = \log{(f_{i}/f_{i-1})} / \log{(\gamma_{i}/\gamma_{i-1})} \,
\end{align}
and introduce the following central-limited slope:
\begin{align}
    p_i = \begin{cases}
        {\rm min} [p^{\rm r}_i,p^{\rm l}_i]\quad {\rm if} \quad p^{\rm r}_i > 0,\\
         {\rm max} [p^{\rm r}_i,p^{\rm l}_i]\quad {\rm if} \quad p^{\rm r}_i < 0,\\
         0 \quad {\rm if} \quad p^{\rm r}_i p^{\rm l}_i \le 0,\\
         p_1 = p_1^{\rm r},\\
         p_N = 0,
    \end{cases}    
\end{align}
such that:
\begin{align}
    f_{i+\frac{1}{2}} &= f_{i} + \frac{p_i}{\gamma_i}f_{i} [\gamma_{i+\frac{1}{2}} - \gamma_i] = f_{i} \left[1 + p_i \left(\frac{\gamma_{i+\frac{1}{2}}}{\gamma_i} -1\right)\right] \nonumber\\
    &= f_i \Delta f^{\rm r}_i\,,\\
    f_{i-\frac {1}{2}} &= f_{i} + \frac{p_i}{\gamma_i}f_{i} [\gamma_{i-\frac{1}{2}} - \gamma_i] = f_{i} \left[1 + p_i \left(\frac{\gamma_{i-\frac{1}{2}}}{\gamma_i} -1\right)\right]\nonumber\\
    &= f_i \Delta f^{\rm l}_i\,,
\end{align}
where we have introduced the relative increments $\Delta f$. A second-order upwind definition of the $\gamma$-interface fluxes is:
\begin{align}
    \dot{\gamma}_{i+\frac{1}{2}}f^n_{i+\frac{1}{2}} =
    \begin{cases}
         \dot{\gamma}_{i+\frac{1}{2}}f^n_i \Delta f^{\rm r}_i\quad{\rm for}\quad \dot{\gamma}_{i+\frac{1}{2}} \ge 0 ,\\
         \dot{\gamma}_{i+\frac{1}{2}}f^n_{i+1}\Delta f^{\rm l}_{i+1} \quad{\rm for}\quad \dot{\gamma}_{i+\frac{1}{2}} \le 0,
    \end{cases}
\end{align}
that simplifies into:
\begin{align}
    \dot{\gamma}_{i+\frac{1}{2}}f^n_{i+\frac{1}{2}} = {\rm Max}[\dot{\gamma}_{i+\frac{1}{2}},0]\Delta f^{\rm r}_i f^n_i + {\rm Min}[\dot{\gamma}_{i+\frac{1}{2}}, 0]\Delta f^{\rm l}_{i+1}f^n_{i+1}\,,  
\end{align} 
and the elements of the tridiagonal matrix to invert are:
\begin{align}
     {\rm A}_{i,i} &= \left[1+\frac{\Delta t}{\Delta\gamma_i} \left({\rm Max}[\dot{\gamma}_{i+\frac{1}{2}},0]\Delta f^{\rm r}_i - {\rm Min}[\dot{\gamma}_{i-\frac{1}{2}}, 0]\Delta f^{\rm l}_i  \right) \right], \\
     {\rm A}_{i,i+1} &= \frac{\Delta t}{\Delta\gamma_i} \left({\rm Min}[\dot{\gamma}_{i+\frac{1}{2}}, 0]\Delta f^{\rm l}_{i+1} \right), \\
     {\rm A}_{i,i-1} &= -\frac{\Delta t}{\Delta\gamma_i} \left( {\rm Max}[\dot{\gamma}_{i-\frac{1}{2}},0]\Delta f^{\rm r}_{i-1}\right). 
\end{align}
This second order scheme improves energy conservation with respect to a fist order scheme by a factor 10.
\section{Discretization of self-synchrotron Compton losses in Klein-Nishina regime}
\label{app:kn}
The most time consuming part in the evolution of the particle distribution function (in general of all the algorithm), is the computation of self-synchotron-Compton losses, especially if done in the full Klein-Nishina regime. For this reason, developing optimized algorithms is quite important. Here we show how to do it in a fully algebric-vectorization way, without the need of interpolations or integrations that can be quite numerically heavy. \citet{Aharonian_Atoyan81a} provide  a formula for the IC losses in the full Klein-Nishina (KN) regime for electrons with Lorentz factor $\gamma$ interacting with a monochromatic radiation field at frequency $\nu$ according to:
\begin{align}
  P_{\rm IC}(\gamma)  = \frac{32}{9}\pi r_{\rm o}^2c\; U_{\rm ph} \gamma^2 \mathcal{G}(\xi) = \frac{4}{3}c \sigma_{\rm T}\mathcal{G}(\xi) U_{\rm ph} \gamma^2\,,
\end{align}
where:
\begin{align}
\mathcal{G}(\xi) &= \frac{9}{\xi^3}\bigg[\left(6+\frac{\xi}{2}+\frac{6}{\xi}\right)\ln{(1+\xi)} +\nonumber\\
&- \ln^2{(1+\xi)} - \frac{(11/12)\xi^3+8\xi^2+13\xi+6}{(1+\xi)^2} +\nonumber\\
&+2\int_{1/(1+\xi)}^1 \frac{\ln{(y)}}{1-y}dy \bigg],
\end{align}
with the IC parameter:
\begin{align}
\xi = 4\gamma \frac{h\nu}{m_e c^2}\,.
\end{align}
Inspired by \citet{Khangulyan_Aharonian+14a}, we can approximate the function $\mathcal{G}(\xi)$ to get an effective cross section:
\begin{align}
\sigma(\xi) = \sigma_{\rm T} \frac{7.65}{\xi}\frac{\ln{\left[1+\frac{\xi}{7.65} \right]}}{\left( 1+1.6505\xi\right)}\,.
\end{align}
Under the mono-chromatic approximation for synchrotron emission, an emitting particle with Lorentz factor $\gamma_i$ will radiate energy at a frequency:
\begin{align}
    \nu_i = 0.29\frac{3e}{4\pi m_e c}B\gamma_i^2 
\end{align}
and for a particle with Lorentz factor $\gamma_j$ the IC parameter will be:
\begin{align}
    \xi_{ij} = \frac{0.87}{\pi}\frac{he}{m_e^2 c^3}B\gamma_i^2 \gamma_j\,,
\end{align}
Now the synchrotron power emitted at frequency $\nu_i$ will be given by:
\begin{align}
    S_{\nu_i} = \frac{4}{3} c \sigma_{\rm T} \frac{B^2}{8\pi}N(\gamma_i) \gamma_i^2\,.
\end{align}
and the related radiation field averaged over the volume of the nebula (see App.~\ref{app:averu}) will be:
\begin{align}
    \langle U_{\nu_i} \rangle = \frac{9}{4}\frac{1}{3} \frac{\sigma_{\rm T}}{\pi R^2} \frac{B^2}{8\pi}N(\gamma_i) \gamma_i^2 = 3\frac{\sigma_{\rm T}}{4\pi R^2} \frac{B^2}{8\pi}N(\gamma_i) \gamma_i^2\,.
\end{align}
It is then possible to define an effective radiation photon field for particles with Lorentz factor $\gamma_j$ as:
\begin{align}
\langle U_{\nu} \rangle_j = \frac{3}{4\pi R^2} \frac{B^2}{8\pi}\sum_{i=1}^n \sigma(\xi_{ij})N(\gamma_i) \gamma_i^2 \,,
\end{align}
such that the energy losses are: 
\begin{align}
P_{\rm IC}^{\rm SSC}(\gamma_j) = R c\sigma_{\rm T}^2\frac{B^2}{2}\sum_{i=1}^n \frac{\sigma(\xi_{ij})}{\sigma_{\rm T}}n(\gamma_i) \gamma_i^2 \gamma_j^2\,.
\end{align}
\section{Average photon field in an optically thin sphere}
\label{app:averu}
A parcel of isotropically emitting plasma, located at $\boldsymbol{x}_{\rm e}$, with volume ${\rm d} \mathcal{V}$ and with mono-chromatic emissivity $\epsilon_\nu$ at frequency $\nu$, will produce an energy flux at a point $\boldsymbol{x}_{\rm o}$ such that the photon density at $\boldsymbol{x}_{\rm o}$ will be given by:
\begin{align}
    &\frac{\epsilon_\nu}{4\pi D^2}{\rm d} \mathcal{V} = n_{\gamma}(\boldsymbol{x}_{\rm o}) h\nu c \nonumber\\
    &\quad \Rightarrow  \quad U_{\gamma}(\boldsymbol{x}_{\rm o}) = n_{\gamma}(\boldsymbol{x}_{\rm o}) h\nu = \frac{\epsilon_\nu}{4\pi c D^2}{\rm d} \mathcal{V}\,,
\end{align}
where $D=\sqrt{(\boldsymbol{x}_{\rm o}-\boldsymbol{x}_{\rm e})^2}$, and $U_\gamma$ is the associated energy density of the radiation field. Let us consider a spherical shell of radius $r$ and thickness ${\rm d} r$, and a point located on the $z$-axis at coordinate $z_{\rm o}$; assuming uniform emissivity, one can compute the total energy density integrating over the shell:
\begin{align}
U_{\gamma}(r,z_{\rm o}) &= 2\pi\int_{0}^\pi \frac{\epsilon_\nu}{4\pi c}\frac{\sin{\theta} {\rm d} \theta}{(r\sin{\theta})^2+(r\cos{\theta} -z_{\rm o})^2}
{\rm d} r\nonumber\\
&= \frac{\epsilon_\nu{\rm d} r}{2 c}\int_{1}^{-1} \frac{-{\rm d} \mu}{r^2+z_{\rm o}^2-2z_{\rm o}r\mu }
\nonumber\\
&= \frac{\epsilon_\nu {\rm d} r}{2 c}\frac{1}{2z_{\rm o} r}\ln{\left[\frac{r^2+z_{\rm o}^2 +2z_{\rm o}r}{r^2+z_{\rm o}^2-2z_{\rm o}r} \right]}\,.
\end{align}
Note that in the limit $z_{\rm o} \rightarrow 0$ one finds $U_{\gamma} = \epsilon_\nu {\rm d} r / r^2$ while for  $z_{\rm o} \rightarrow \infty$ one has $U_{\gamma} = \epsilon_\nu {\rm d} r / z_{\rm o}^2$. 
Integrating over a spherical volume extending from $r=0$ to $r=R$ one gets the total radiation field at $z_{\rm o}$:
\begin{align}
U_{\gamma}(z_{\rm o}) & = \frac{\epsilon_\nu }{4 c z_{\rm o}}\int_0^R r\ln{\left[\frac{r^2+z_{\rm o}^2 +2z_{\rm o}r}{r^2+z_{\rm o}^2-2z_{\rm o}r} \right]}{\rm d} r\,.
\end{align}
Now we are interested in the illumination of points inside $R$. In the limit $z_{\rm o} \rightarrow 0$  one finds $U_{\gamma}(0) = \epsilon_\nu R/c $. 
On the other hand, for $z_{\rm o} \rightarrow R$ one has $U_{\gamma}(R) = \epsilon_\nu R/(2c)$. 
We can define an average radiation field  if one integrates over $z_{\rm o}$ within a sphere of radius $R$ to get:
\begin{align}
\langle &U_{\gamma}\rangle = \frac{3}{4\pi R^3}\int_0^R 4\pi U_{\gamma}(z_{\rm o})z_{\rm o}^2 {\rm d}z_{\rm o} = \frac{3}{4}\frac{R}{c}\epsilon_\nu \\
&= {\rm Re}\left[\frac{\epsilon_\nu z_{\rm o}}{4 c } \left\{\frac{2R}{z_{\rm o}}-\!2{\rm ArcTanh}\left(\frac{R}{z_{\rm o}} \right) +\frac{R^2}{2z_{\rm o}^2}\ln{\left[ \frac{(R+z_{\rm o})^2}{(R-z_{\rm o})^2}\right] }\right\}\right]\,.
\end{align}
Now, in an optically thin source, one defines the emissivity based on the total emitted power $\mathcal{S}_\nu$ as: $\mathcal{S}_\nu = 4\pi \epsilon_\nu R^3 /3$. This leads to the following relation:
\begin{align}
\langle U_{\gamma}\rangle = \frac{9}{4}\frac{\mathcal{S}_\nu}{4\pi R^2c}\,.
\end{align}

\end{appendix}
\end{document}